\documentclass[sigconf]{acmart}
\AtBeginDocument{%
  }

\copyrightyear{2026}
\acmYear{2026}
\setcopyright{cc}
\setcctype{by}
\acmConference[KDD '26]{Proceedings of the 32nd ACM SIGKDD Conference on Knowledge Discovery and Data Mining V.1}{August 09--13, 2026}{Jeju Island, Republic of Korea}
\acmBooktitle{Proceedings of the 32nd ACM SIGKDD Conference on Knowledge Discovery and Data Mining V.1 (KDD '26), August 09--13, 2026, Jeju Island, Republic of Korea}
\acmPrice{}
\acmDOI{10.1145/3770854.3780176}
\acmISBN{979-8-4007-2258-5/2026/08}

\usepackage{tikz}
\usepackage{amsmath}

\usepackage[ruled,linesnumbered]{algorithm2e}

\usepackage{amsfonts}

\usepackage{graphicx}
\usepackage{subfigure}

\usepackage{minted}

\usepackage{multirow}

\usepackage{booktabs}
\usepackage{makecell}

\usepackage{enumitem}

\usepackage{siunitx}

\usepackage[normalem]{ulem}

\usepackage[utf8]{inputenc}

\usepackage{cleveref}
\crefname{section}{§}{§§}

\usepackage[table,xcdraw]{xcolor}

\usepackage[font=normalsize,skip=4pt]{caption}
\newcommand{\distance}{4pt}
\usepackage{grumble}

\begin{document}

\title[FuXi-$\gamma$]{FuXi-$\gamma$: Efficient Sequential Recommendation with Exponential-Power Temporal Encoder and \\Diagonal-Sparse Positional Mechanism}

\author{Dezhi Yi}
\email{dezhi.yi@mail.nankai.edu.cn}
\affiliation{%
  \department{College of Computer Science, DISSec}
  \institution{Nankai University}
  \city{Tianjin}
  \country{China}}

\author{Wei Guo}
\email{guowei67@huawei.com}
\affiliation{%
  \institution{Huawei Technologies}
  \city{Shanghai}
  \country{China}}

\author{Wenyang Cui}
\email{cuiwenyang@mail.nankai.edu.cn}
\affiliation{%
  \department{College of Computer Science, DISSec}
  \institution{Nankai University}
  \city{Tianjin}
  \country{China}}

\author{Wenxuan He}
\email{wxhe@mail.nankai.edu.cn}
\affiliation{%
  \department{College of Computer Science, DISSec}
  \institution{Nankai University}
  \city{Tianjin}
  \country{China}}

\author{Huifeng Guo}
\email{huifeng.guo@huawei.com}
\affiliation{%
  \institution{Huawei Technologies}
  \city{Shenzhen}
  \country{China}}

\author{Yong Liu}
\email{liu.yong6@huawei.com}
\affiliation{%
  \institution{Huawei Technologies}
  \city{Shanghai}
  \country{China}}

\author{Zhenhua Dong}
\email{dongzhenhua@huawei.com}
\affiliation{%
  \institution{Huawei Technologies}
  \city{Shenzhen}
  \country{China}}

\author{Ye Lu}
\authornote{Corresponding author.}
\email{luye@nankai.edu.cn}
\affiliation{%
  \department{College of Computer Science, DISSec}
  \institution{Nankai University}
  \city{Tianjin}
  \country{China}}

\renewcommand{\shortauthors}{Dezhi Yi et al.}

\begin{abstract}

Sequential recommendation aims to model users’ evolving preferences based on their historical interactions. 
Recent advances leverage Transformer-based architectures to capture global dependencies, but existing methods often suffer from high computational overhead, primarily due to discontinuous memory access in temporal encoding and dense attention over long sequences. 
To address these limitations, we propose FuXi-$\gamma$, a novel sequential recommendation framework that improves both effectiveness and efficiency through principled architectural design. 
FuXi-$\gamma$ adopts a decoder-only Transformer structure and introduces two key innovations: 
(1) An exponential-power temporal encoder that encodes relative temporal intervals using a tunable exponential decay function inspired by the Ebbinghaus forgetting curve. 
This encoder enables flexible modeling of both short-term and long-term preferences while maintaining high efficiency through continuous memory access and pure matrix operations. 
(2) A diagonal-sparse positional mechanism that prunes low-contribution attention blocks using a diagonal-sliding strategy guided by the persymmetry of Toeplitz matrix. 
Extensive experiments on four real-world datasets demonstrate that FuXi-$\gamma$ achieves state-of-the-art performance in recommendation quality, while accelerating training by up to \textbf{4.74$\times$} and inference by up to \textbf{6.18$\times$}, making it a practical and scalable solution for long-sequence recommendation. 
Our code is available at \url{https://github.com/Yeedzhi/FuXi-gamma}. 

\end{abstract}




\maketitle

\section{Introduction}
Sequential recommendation aims to predict users’ next interaction based on their historical behavior sequences~\cite{kang2018self, sun2019bert4rec, ye2025fuxi}. 
Accurate predictions improve user experience and drive substantial commercial value~\cite{kaasinen2009user, liu2020category, wang2020survey}. 
The training efficiency of recommendation models affects their update frequency and service freshness, while inference efficiency impacts system latency and deployment cost~\cite{li2023ctrl, wang2023bert4ctr, li2023fragment}. 
Hence, designing recommendation frameworks that are both effective and efficient is a core challenge.

Recent advances in generative recommendation models~\cite{chen2024hllm, zhai2024actions, ye2025fuxi} leverage autoregressive architectures to achieve superior performance and scaling effects compared to traditional methods~\cite{kang2018self, hidasi2015session, sun2019bert4rec}. 
However, these models often introduce efficiency bottlenecks due to architectural complexity. 
For example, although HSTU~\cite{zhai2024actions} improves HR@10 by 55.18\% over SASRec~\cite{kang2018self}, it suffers from a 3.21$\times$ slowdown in training efficiency. 
FuXi-$\alpha$~\cite{ye2025fuxi} further improves HR@10 by 5.66\% over HSTU but incurs an additional 9.63\% efficiency loss. 
Such inefficiency forces production systems to reduce model size to meet latency requirements, weakening the scaling benefits of the autoregressive paradigm.

A primary efficiency bottleneck in sequential recommendation lies in temporal encoding module. 
Temporal information is essential for capturing the dynamics of user preferences over time, which is fundamental to modeling interest patterns~\cite{li2020time, ye2020time, cho2020meantime}. 
However, temporal encoders in existing state-of-the-art frameworks often conflict with parallel hardware due to irregular and fragmented memory access, severely limiting computational efficiency. 
For example, both HSTU~\cite{zhai2024actions} and FuXi-$\alpha$~\cite{ye2025fuxi} employ a T5-style bucket-based encoding scheme~\cite{raffel2020exploring}, where relative temporal intervals are log-transformed and discretized into distinct indices for bucket lookup. 
This operation pattern results in $n^2$ non-contiguous and unstructured memory accesses for a sequence of length $n$, posing a major obstacle to parallel execution on modern hardware. 
Although FuXi-$\beta$~\cite{ye2025beta} introduces an inverse-proportion decay function to alleviate some of these issues, both approaches lack grounding in cognitive principles, limiting their expressiveness and explainability.

Another overlooked inefficiency stems from positional encoding module. 
Modern frameworks introduce relative positional encoding even when absolute positional embedding and temporal signal are already available. 
For example, FuXi-$\alpha$~\cite{ye2025fuxi} and FuXi-$\beta$~\cite{ye2025beta} incorporate positional encoding as a separate attention channel with $O(n^2)$ complexity, which becomes an efficiency bottleneck for long sequences. 
Recent advances in block-sparse semantic attention~\cite{gao2024seerattention, jiang2024minference, lai2025flexprefill, xu2025xattention} offer promising solutions, but have yet to be explored in the context of positional attention in recommendation.

To address dual challenges of efficiency and effectiveness in modeling user interests, we propose a novel sequential recommendation framework named FuXi-$\gamma$. 
Building upon FuXi-$\beta$~\cite{ye2025beta}, FuXi-$\gamma$ adopts an autoregressive architecture with dual attention channels for temporal and positional modeling. 
We introduce two key innovations:
(1) An exponential-power temporal encoder, inspired by the Ebbinghaus forgetting curve~\cite{ebbinghaus1885memory}, which models time-aware user interest decay through a continuous, fully matrix-based formulation. 
This cognitively motivated design offers strong expressiveness and hardware-friendly execution, yielding up to \textbf{11.00$\times$} speedup over the bucket-based encoder and significant gains in recommendation performance. 
(2) A diagonal-sparse positional mechanism, which identifies and prunes low-contribution attention blocks based on importance scoring on the leftmost column, reducing positional attention overhead by \textbf{74.56\%} while preserving recommendation quality. 
Together with practical optimizations such as data type pre-conversion, FuXi-$\gamma$ achieves up to \textbf{4.74$\times$} and \textbf{6.18$\times$} improvements in training and inference efficiency, respectively, while consistently outperforming state-of-the-art baselines across multiple datasets.

Our main contributions are summarized as follows:

\begin{itemize}[leftmargin=1.2em]

\item
We propose a novel sequential recommendation framework, FuXi-$\gamma$, that enhances both effectiveness and efficiency, particularly in scenarios involving long sequences. 

\item
We design an exponential-power temporal encoder that models user interest decay through a tunable exponential function. 
It offers flexibility to adapt to diverse behavioral patterns and achieves high efficiency through regular memory access and pure matrix operations. 

\item
We introduce a diagonal-sparse positional mechanism that prunes redundant attention blocks, substantially reducing computational overhead while preserving recommendation accuracy. 

\item
We conduct extensive experiments on four real-world datasets, demonstrating that FuXi-$\gamma$ achieves state-of-the-art performance. 
On a large-scale industrial music dataset, FuXi-$\gamma$ achieves improvements of \textbf{25.06\%} in HR@10 and \textbf{42.86\%} in NDCG@10 over strong autoregressive baselines. 

\end{itemize}

\section{Related Work}

\subsection{Sequential Recommendation}
Sequential recommendation aims to predict users’ future preferences based on their interaction sequences. 
Early methods employ probabilistic models, such as Markov Chains~\cite{rendle2010factorizing} and session-based kNN~\cite{he2016fusing, hu2020modeling}, to capture short-term transitions. 
Subsequent deep learning approaches model higher-order dependencies through CNNs~\cite{tang2018personalized}, RNNs~\cite{liu2016context}, and GRU-based architecture~\cite{hidasi2015session}. 
With the introduction of Transformer~\cite{vaswani2017attention}, global dependency modeling becomes possible via self-attention. 
SASRec~\cite{kang2018self} uses multi-head self-attention to identify important items. 
BERT4Rec~\cite{sun2019bert4rec} leverages bidirectional encoding for improved prediction. 
Further extensions explore more complex settings, including cross-domain~\cite{zhao2023sequential}, multi-modal~\cite{zhang2024multimodal}, and multi-behavior scenarios~\cite{su2023personalized}. 

Traditional frameworks typically follow a ranking-based formulation, which limits their scalability and adaptability. 
Generative recommendation reframes the task as sequence generation, enabling unified, end-to-end modeling. 
TIGER~\cite{rajput2023recommender} pioneers generative recommendation for zero-shot scenarios. 
HLLM~\cite{chen2024hllm} tokenizes item IDs for autoregressive modeling with large language models. 
HSTU~\cite{zhai2024actions} scales recommendation using Transformer-based generation. 
FuXi-$\alpha$~\cite{ye2025fuxi} introduces a three-channel design to separately model semantic, temporal, and positional features, while FuXi-$\beta$~\cite{ye2025beta} improves efficiency by removing semantic modeling. 

Despite recent progress, balancing effectiveness and efficiency remains a core challenge. 
To this end, we propose FuXi-$\gamma$ that integrates a cognitively inspired temporal encoder and a sparse positional mechanism to improve both performance and scalability.

\subsection{Temporal Encoder}

Temporal information plays a crucial role in capturing the evolving nature of user interests. 
Prior works explore various strategies, such as encoder-decoder timestamp modeling~\cite{zhang2023time}, joint learning of temporal patterns~\cite{ye2020time}, and multi-granularity temporal embeddings~\cite{cho2020meantime}. 
TiSASRec~\cite{li2020time} explicitly incorporates both absolute positions and relative temporal intervals. 
Time2Vec~\cite{kazemi2019time2vec} represents temporal signals through sinusoidal and linear activations. 
More recently, generative recommendation models like HSTU~\cite{zhai2024actions} and FuXi-$\alpha$~\cite{ye2025fuxi} adopt T5-style bucket encodings~\cite{raffel2020exploring} to discretize relative temporal intervals. 
Despite effectiveness, the bucket-based method introduces discontinuous memory access and poor hardware utilization, resulting in substantial efficiency bottlenecks, particularly for long sequences. 
FuXi-$\beta$~\cite{ye2025beta} replaces bucketing with an inverse-proportional decay function, offering improved efficiency. 
However, its fixed decay pattern lacks adaptability across different application scenarios, limiting modeling flexibility. 

To overcome these limitations, we propose an exponential-power temporal encoder inspired by the Ebbinghaus forgetting curve~\cite{ebbinghaus1885memory}. 
The design enables flexible modeling of temporal decay patterns and maintains efficient, hardware-friendly continuous memory access.

\subsection{Positional Encoder}

Positional encoding provides contextual information for sequence modeling tasks. 
Existing approaches can be broadly classified into absolute and relative encoding methods. 
Absolute encodings directly incorporate positional vectors into item representations, typically implemented using either fixed sinusoidal functions~\cite{vaswani2017attention} or learnable embeddings~\cite{gehring2017convolutional}. 
Relative encodings~\cite{shaw2018self, dai2019transformer, he2020deberta, huang2020improve, ke2020rethinking, raffel2020exploring} model pairwise item distances by learning weights associated with relative positional offsets. 
Recent generative recommendation models often combine both encoding types, integrating relative encoding either as an additive bias in the semantic channel~\cite{zhai2024actions} or as a separate attention matrix in an independent channel~\cite{ye2025fuxi, ye2025beta}.

Following FuXi-$\alpha$~\cite{ye2025fuxi} and FuXi-$\beta$~\cite{ye2025beta}, we adopt relative positional encoding as a separate attention matrix. 
However, we observe that positional information is already partially captured by the embedding layer and the temporal encoder, introducing redundancy. 
To mitigate this, we design a diagonal-sparse pruning method applied after training. 
This method removes low-contribution blocks from the positional attention matrix, significantly reducing computation overhead while preserving recommendation accuracy. 

\section{Methodology}

In this section, we begin by formally defining the next-item sequential recommendation task. 
We then present the overall architecture of FuXi-$\gamma$, followed by a detailed description of its core components.

\subsection{Problem Statement}
\label{sec: problem statement}
The primary objective of sequential recommendation is to predict the next item a user is likely to interact with, conditioned on their historical interaction sequence. 
Formally, let $\mathcal{U} = \left\{ u_{1}, u_{2}, \ldots, u_{\left| \mathcal{U} \right|} \right\}$ denote the set of users, and $\mathcal{V} = \left\{ v_{1}, v_{2}, \ldots, v_{\left| \mathcal{V} \right|} \right\}$ the set of items. 
For each user $u \in \mathcal{U}$, the interaction history is represented as a time-ordered sequence $S_{u} = \left[ ( v_{1}^{\left( u \right)}, t_{1}^{\left( u \right)} ), ( v_{2}^{\left( u \right)}, t_{2}^{\left( u \right)} ), \ldots, ( v_{n_{u}}^{\left( u \right)}, t_{n_{u}}^{\left( u \right)} ) \right] $, where each tuple $ ( v_i^{\left( u \right)}, t_i^{\left( u \right)} ) $ corresponds to an item interacted with at timestamp $t_i^{\left( u \right)}$. 
The goal is to predict the subsequent item $v_{n_u+1}^{\left( u \right)}$ that user $u$ will interact with, given their historical sequence $S_u$. 
This task is typically formulated as estimating the conditional probability distribution over the item space: $P\left(v_{n_{u}+1}^{\left( u \right)}=v \mid S_{u}\right), \forall v \in \mathcal{V}$. 
During training, the model is optimized to predict the next item $v_{i+1}^{\left( u \right)}$ at each position $i$ along the sequence $S_u$. 
Thus the desired output sequence corresponds to $ \left[ v_{2}^{\left( u \right)}, v_{3}^{\left( u \right)}, \ldots, v_{n_{u}+1}^{\left( u \right)} \right] $~\cite{kang2018self}.

\begin{figure}
\centering
  \includegraphics[width=0.6\linewidth]{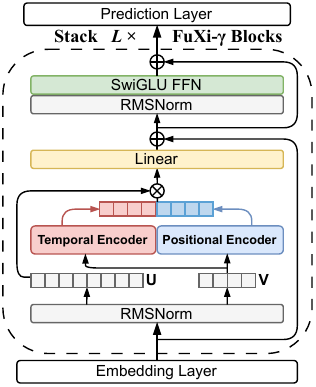}
  \caption{Overall architecture of FuXi-$\gamma$.}
  \label{fig: overview}
\end{figure}

\subsection{Overall Architecture}
\label{sec: overall architecture}
The overall architecture of FuXi-$\gamma$ is illustrated in Figure~\ref{fig: overview}. 
Following the design of FuXi-$\beta$~\cite{ye2025beta}, FuXi-$\gamma$ adopts a decoder-only Transformer architecture without query-key attention, aiming to streamline computation while preserving modeling capacity. 
Specifically, FuXi-$\gamma$ consists of three components: 
(1) Embedding Layer: Encodes input items with learned item embeddings and incorporates absolute positional information.
(2) FuXi-$\gamma$ Block: Capture user interest patterns via dual-channel self-attention layer and SwiGLU FFN.
(3) Prediction Layer: Computes the probability distribution over the item vocabulary for the next-item prediction.

\subsubsection{Embedding Layer}
To address the variability in user interaction sequence lengths, we normalize each user’s interaction history to a fixed length $n$ prior to the embedding layer. 
Sequences longer than $n$ are truncated, while shorter sequences are padded with a special "padding item". 
Each item $v \in \mathcal{V}$ is mapped to a $d$-dimensional vector via a learnable embedding matrix $E \in \mathbb{R}^{\left|\mathcal{V}\right| \times d}$. 
In addition, we incorporate absolute positional information by adding learnable positional embeddings. 
Let $e_i$ and $p_i$ denote the item embedding and positional embedding of the $i$-th position, respectively.
For a user sequence $S_u$, the output of the embedding layer is given by:
\begin{equation}
X^{0} = \left[ e_{1}^{(u)} + p_{1}, \ldots, e_{n_u}^{(u)} + p_{n_u}, \mathbf{0}, \ldots, \mathbf{0} \right],
\label{eq: embedding layer}
\end{equation}
where vectors $\mathbf{0}$ correspond to padding positions from $n_u + 1$ to $n$.

The resulting embedding $X^{0}$ is subsequently passed through a stack of FuXi-$\gamma$ blocks to capture user interest patterns. 
The design details of FuXi-$\gamma$ block are described in Section~\ref{sec: fuxi block}.

\subsubsection{Prediction Layer \& Optimization Objective}
After passing through $L$ stacked FuXi-$\gamma$ blocks, each position in the sequence encodes sufficient contextual information from previously interacted items. 
To generate the prediction, we project the final hidden representation onto the item space by computing dot product with the transpose of input embedding matrix, followed by a softmax function to obtain a probability distribution over all candidate items: 
\begin{equation}
P\left(v_{i}^{(u)} = v \mid \left[ (v_{1}^{(u)}, t_{1}^{(u)}), \ldots, (v_{i-1}^{(u)}, t_{i-1}^{(u)}) \right] \right) = \mathrm{Softmax} \left( x^{L} E^\mathsf{T} \right)_{v},
\label{eq: prediction layer}
\end{equation}
where $x^{L}$ denotes the final-layer representation at position $i-1$, and $E$ is the input embedding matrix. 
To improve training efficiency, we adopt the sampled softmax loss, where the softmax is computed over the true item and $N$ randomly sampled negative items~\cite{klenitskiy2023turning}.

\subsection{FuXi-$\gamma$ Block}
\label{sec: fuxi block}

FuXi-$\gamma$ block comprises two components: a dual-channel self-attention layer and a SwiGLU feed-forward network (FFN)~\cite{shazeer2020glu}. 
To improve training stability, we adopt a pre-normalization~\cite{grattafiori2024llama} strategy, applying layer normalization before each sub-layer computation.

\subsubsection{Dual-Channel Self-Attention Layer}
\label{sec: dual-channel self-attention}
This layer decouples representation learning into two distinct channels: a temporal channel and a positional channel. 
This design enables each channel's encoder to specialize in capturing different aspects of the sequence information. 
Then, we adopt a post-mixing strategy via point-wise interaction and linear projection to aggregate signals.

Let $X \in \mathbb{R}^{n \times d}$ denote the input embedding to one FuXi-$\gamma$ block. 
We first construct two projection representations of input sequence:
\begin{equation}
U,\ V = \mathrm{Split}(\phi(\mathrm{RMSNorm}(X)W_{uv})),
\label{eq: initial layer}
\end{equation}
where RMSNorm denotes root mean square layer normalization~\cite{zhang2019root}, 
$W_{uv} \in \mathbb{R}^{d \times 3d}$ is a learnable matrix, 
$\phi$ denotes SiLU nonlinear transformation~\cite{elfwing2018sigmoid}, 
$U \in \mathbb{R}^{n \times 2d}$ and $V \in \mathbb{R}^{n \times d}$ are two projected outputs. 

To simplify computation and improve efficiency, we share $V$ between temporal and positional encoders to learn user interests. 
Then the outputs of two encoders are concatenated and mixed with $U$ via Hadamard product to enable explicit 2-order interactions: 
\begin{equation}
I = \mathrm{RMSNorm}(\mathrm{Concat}(A_{ts}V, W_{pos}V)) \odot U,
\label{eq: interaction layer}
\end{equation}
where $A_{ts} \in \mathbb{R}^{n \times n}$ is a temporal attention matrix constructed based on relative temporal intervals (see Section~\ref{sec: design of temporal encoder} for design details); 
$W_{pos} \in \mathbb{R}^{n \times n}$ is a learnable positional encoding matrix, where $W_{pos}^{i,j}$ denotes the attention weight for relative positional offset $i-j$ and satisfies $W_{pos}^{i,j} = W_{pos}^{i+m,j+m}$, i.e., $W_{pos}$ is a Toeplitz matrix. 

Finally, we pass the interaction result $I \in \mathbb{R}^{n \times 2d}$ through a linear layer to aggregate temporal and positional signals, and introduce a residual connection~\cite{he2016deep} to maintain original sequence information:
\begin{equation}
O = IW_{o} + b + X,
\label{eq: linear projection layer}
\end{equation}
where $W_{o} \in \mathbb{R}^{2d \times d}$ and $b \in \mathbb{R}^{d}$ are learnable parameters.

\subsubsection{SwiGLU FFN}
Each dual-channel self-attention layer is followed by a SwiGLU FFN, which further refines the representation through implicit interactions. 
To preserve the flow of gradients and enable deep modeling, we incorporate a residual connection around the FFN. 
The computations are formally defined as:
\begin{equation}
\begin{split}
&O' = \mathrm{RMSNorm}(O), \\
&H = (\phi (O'W_{1}) \odot (O'W_{2}))W_{3} + O,
\end{split}
\label{eq: swiglu ffn}
\end{equation}
where $\phi$ is SiLU activation, 
$W_{1}, W_{2} \in \mathbb{R}^{d \times d_{\text{FFN}}}$ and $W_{3} \in \mathbb{R}^{d_{\text{FFN}} \times d}$ are learnable weights, 
$H \in \mathbb{R}^{n \times d}$ is the final output of FuXi-$\gamma$ block.

\subsection{Exponential-Power Temporal Encoder}
\label{sec: temporal encoder}

Effectively and efficiently capturing the evolving nature of user interests over time is a key challenge. 
To address this problem, we propose an exponential-power temporal encoder that utilizes a continuous and tunable decay mechanism, offering both strong modeling capabilities and high computational efficiency.

\subsubsection{Algorithm Design}
\label{sec: design of temporal encoder}

Motivated by the intuition that recent interactions better reflect users' current interests, we impose a monotonic decay pattern on temporal attention weights: smaller temporal intervals receive higher weights, while larger intervals receive lower weights. 
Inspired by the Ebbinghaus forgetting curve~\cite{ebbinghaus1885memory}, which describes human memory decays exponentially over time, we adopt a tunable exponential function to model the temporal decay in user preferences. 
First, we represent temporal intervals within a user’s interaction sequence as pairwise differences between item timestamps. 
Specifically, we construct a relative temporal matrix $T \in \mathbb{N}^{n \times n}$ for each sequence, where $T^{i,j}$ denotes the absolute difference between the timestamps of the $i$-th and $j$-th items: $T^{i,j} = |t_i - t_j|$. 
To reduce computation and memory cost, this preprocessing is performed only once, and the resulting $T$ is shared across all temporal encoders. 
Then, we apply a nonlinear transformation to $T$ using a power function. 
This design prevents insufficient learning of long-term user preferences that can occur when excessively long temporal intervals produce very weak signals. 
Finally, we adopt an exponential function to model the decay of user interests over time. 
Formally, the temporal attention matrix $A_{ts} \in \mathbb{R}^{n \times n}$ used in Section~\ref{sec: dual-channel self-attention} is defined as: 
\begin{equation}
A_{ts} = \alpha \cdot \gamma^{T^\beta},\quad \text{i.e.,}\quad A_{ts}^{i,j} = \alpha \cdot \gamma^{|t_i - t_j|^\beta},\quad \gamma \in (0, 1),
\label{eq: temporal encoder}
\end{equation}
where $\alpha \in \mathbb{R}$ is a learnable parameter representing the base interest intensity, $\beta \in \mathbb{R}$ is a learnable parameter controlling the nonlinearity of scaling transformation, and $\gamma \in (0, 1)$ is a decay parameter that governs the rate of interest attenuation. 
A smaller $\gamma$ induces faster decay, emphasizing short-term patterns, whereas a larger $\gamma$ preserves more long-term signals. 
This design offers flexibility across diverse user behaviors and application scenarios, enabling the model to capture both short-term and long-term preferences in a simple yet expressive manner.

\subsubsection{Pre-Conversion of Data Type}
In Equation~\ref{eq: temporal encoder}, the parameters $\alpha$, $\beta$, and $\gamma$ are represented in float32, while the relative temporal matrix $T$ is originally in int64. 
Due to the inconsistency of data types, computing $T^\beta$ implicitly introduces type casting at runtime, which incurs additional time overhead and becomes more costly when repeated across multiple layers. 
To mitigate this issue, we explicitly cast $T$ to float32 during preprocessing. 
This pre-conversion ensures that all subsequent operations, such as power and exponential calculations, are performed using the same data type, thereby avoiding runtime casting and improving computational efficiency. 
This simple optimization brings notable performance gains: 
(1) The execution time of our exponential-power temporal encoder is reduced by 64.82\%. 
(2) Overall FuXi-$\gamma$ achieves an additional 12.61\% speedup and 5.08\% reduction in memory usage during training, and a 15.53\% speedup and 6.98\% memory reduction during inference.

\subsubsection{Analysis}
Our proposed exponential-power temporal encoder achieves strong performance in both architectural compatibility and computational efficiency (see Section~\ref{sec: in-depth study of temporal encoder} for details). 
Existing bucket-based method, as used in HSTU~\cite{zhai2024actions} and FuXi-$\alpha$~\cite{ye2025fuxi}, applies logarithmic transformations to temporal intervals and then maps them into discrete indices for bucket lookup. 
Although this approach offers some modeling flexibility, it suffers from several key limitations: 
(1) The design does not reflect how user interests evolve over time. 
(2) Coarse-grained discretization leads to a loss of detail, as different temporal intervals may fall into the same bucket. 
(3) Most importantly, the irregular memory access during bucket lookup creates a critical efficiency bottleneck, especially when handling long sequences. 
FuXi-$\beta$~\cite{ye2025beta} proposes an inverse-proportion function to represent temporal decay, which helps reduce memory access overhead and partly captures interest attenuation. 
However, it lacks adaptability due to the untunable decay pattern, which overly favors recent interactions and limits flexibility in modeling longer-term user preferences. 
In contrast, our exponential-power encoder is inspired by cognitive memory theory and operates entirely through standard matrix computations. 
This design enables more accurate modeling of adaptive-range interests and ensures smooth and efficient execution on modern hardware.

\subsection{Diagonal-Sparse Positional Mechanism}

Positional information is explicitly encoded in the embedding layer and implicitly reflected through temporal signals. 
Therefore, we identify and prune redundant weights in relative positional attention, further enhancing the inference performance of FuXi-$\gamma$. 
Specifically, we design the diagonal-sparse positional mechanism, which consists of three steps:
(1) Block-level division of the positional attention map. 
(2) Importance scoring of each block. 
(3) Selection of important blocks for sparse attention computation. 

\begin{figure}[H]
\centering
  \includegraphics[width=1.0\linewidth]{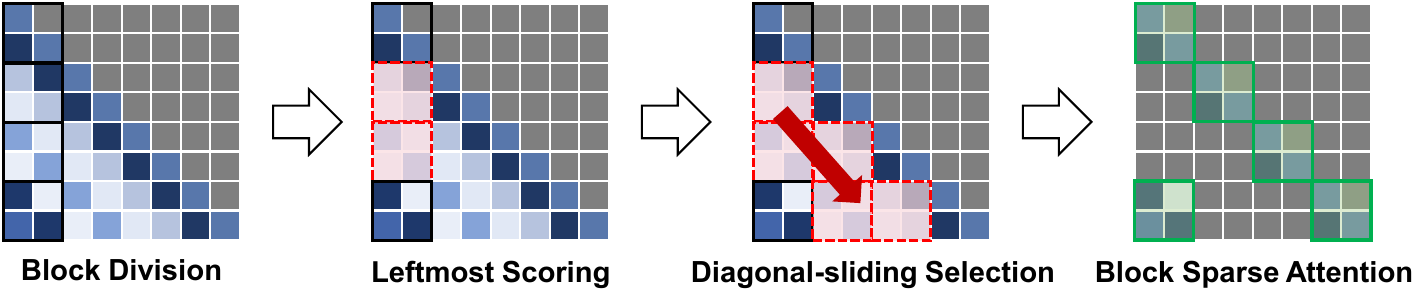}
  \caption{Illustration of diagonal-sparse positional mechanism. In this example, sequence length $n=8$, stride size $s=2$, and configured pruning ratio $\tau=50\%$. Red Blocks are pruned due to lower importance scores. Only the remaining green blocks participate in positional attention computation.}
  \label{fig: sparse scoring strategy}
\end{figure}

\subsubsection{Block Division}
Element-level unstructured pruning offers accuracy advantages but incurs considerable overhead in hardware deployment. 
Row-level and column-level structured pruning are more hardware-friendly but may overlook critical sparse patterns, leading to degraded accuracy~\cite{xie2024winols}. 
To balance trade-off, we adopt block-level semi-structured pruning strategy to retain accuracy and improve deployability. 
Specifically, we partition the positional attention map into several blocks of size $s \times s$, where $s$ is configurable stride size. 
To accommodate sequences whose length $n$ can not be divisible by $s$, we apply zero-padding to the top and right edges of attention map until the dimension becomes divisible. 
Since FuXi-$\gamma$ adopts a decoder-only architecture, the upper-triangular region of the attention map is masked with 0. 
Therefore, our padding operation does not damage the original meaningful information.

\subsubsection{Leftmost Importance Scoring}
An effective scoring method should robustly identify the high-utility regions of attention map while maintaining computational efficiency. 
To this end, we propose the leftmost scoring method as shown in Figure~\ref{fig: sparse scoring strategy}. 
Given that attention map reflects contribution through weighted summation, we use the sum of absolute values within each block as a proxy for importance. 
To reduce computation, we utilize the persymmetry of Toeplitz matrix: $W_{pos}^{i,j} = W_{pos}^{i+m,j+m}$, as established in Section~\ref{sec: dual-channel self-attention}. 
This means all blocks lying on the same diagonal share identical values. 
Consequently, it suffices to compute importance scores only for the blocks in the leftmost column, each of which uniquely represents one diagonal slice of the entire map. 
Our scoring method is effective for two key reasons: 
(1) Coverage Guarantee: Since each positional weight is included in at least one leftmost block, the scoring process ensures full coverage of attention weights. 
(2) Structural Awareness: The leftmost column intersects with all slash-like patterns (i.e., diagonals), enabling efficient detection of meaningful structures and guiding sparse attention accordingly.

\subsubsection{Diagonal-Sliding Selection}
Based on importance scores, we propose a diagonal-sliding block selection method to generate the final sparse attention mask. 
The formal algorithm is summarized in Algorithm~\ref{alg: alg1}. 
Let $n$ denote the sequence length, $s$ the configured stride size, and $\tau \in [0,1]$ the configured pruning ratio. 
After estimating the importance scores of $\frac{n}{s}$ leftmost blocks, we identify the top-$k$ most unimportant blocks, where $k = \left \lfloor \frac{n}{s} \cdot \tau \right \rfloor$. 
Then, we slide these identified blocks diagonally across the entire attention map to generate final sparse mask. 
This overall sparsity mechanism effectively and efficiently identifies and prunes redundant blocks within the positional attention map in a diagonally structured manner.

\begin{algorithm}
    \footnotesize
 \setcounter{AlgoLine}{0}
 \caption{{Diagonal-Sliding Sparse Mask Generation}}
 \label{alg: alg1}
 \DontPrintSemicolon
 \SetCommentSty{mycommfont}
 {
    \KwIn{Positional attention map $W_{pos} \in \mathbb{R}^{n \times n}$; Stride size $s$; Pruning ratio $\tau$}
    \KwOut{Sparse mask $M$}
    
    {\color[HTML]{E66D68} \textbf{Block Division:}} \\
    $num\_blocks \gets \frac{n}{s}$ \\
    $leftmost\_blocks \gets W_{pos}[:, :s].{\sf{view}}(num\_blocks, s, s)$ \\
    
    {\color[HTML]{E66D68} \textbf{Leftmost Importance Scoring:}} \\
    $abs\_sum \gets {\sf{sum}}(|leftmost\_blocks|, \text{dim} = (2, 3))$ \\
    $max\_index \gets num\_blocks \times num\_blocks - 1$ \\
    $num\_mask \gets \lfloor num\_blocks \times \tau \rfloor$ \\
    $\_, indices\_to\_mask \gets {\sf{TopK}}(abs\_sum, \text{k} = num\_mask, \text{largest} = \text{False})$ \\
    $real\_indices \gets indices\_to\_mask \times num\_blocks$ \\
    
     {\color[HTML]{E66D68} \textbf{Diagonal-Sliding Selection:}} \\
    $M \gets [\,]$ \\
    $increment \gets num\_blocks + 1$ \\
    \ForEach{$index \in real\_indices$}{
        \While{$index \le max\_index$}{
            $M.{\sf{append}}(index)$ \\
            $index \gets index + increment$
        }
    }
    
    \Return{$M$}
}
\end{algorithm}

\section{Experiment}

\subsection{Experimental Setup}

\begin{table}[H]
\centering
\caption{Statistics of the datasets.}
\label{table: statistics of the datasets}
\scalebox{0.88}{
\begin{tabular}{c|cccc}
\Xhline{2\arrayrulewidth}
\textbf{Datasets} & \textbf{\#Users}             & \textbf{\#Items}             & \textbf{\#Actions}               & \textbf{Avg. Len.}           \\ \hline
ML-1M             & 6,040                        & 3,706                        & 1,000,209                        & 165.60                       \\
ML-20M            & 138,493                      & 26,744                       & 20,000,263                       & 144.41                       \\
KuaiRand          & 25,634                       & 7,550                        & 1,432,897                        & 55.90                       \\
Industrial        & 28,927,689                   & 476,544                      & 1,313,729,225                    & 45.41                        \\ \Xhline{2\arrayrulewidth}
\end{tabular}
}
\end{table}

\begin{table*}[]
\centering
\caption{Overall recommendation performance comparison on the public datasets. The best result is highlighted in bold, and the second-best result is underlined. Superscript * indicates the statistically significant improvements (i.e., two-sided t-test with $p$ < 0.05). NG@10 and NG@50 stand for NDCG@10 and NDCG@50, respectively. Infer. denotes the inference latency: we normalize our FuXi-$\gamma$’s latency to 1.00, and report the latencies of all other models as multiples relative to FuXi-$\gamma$.}
\label{table: overall recommendation performance comparison on the public datasets}
\scalebox{0.75}{
\setlength{\tabcolsep}{3pt}
\begin{tabular}{c|ccccc|cccccc|ccccc}
\Xhline{2\arrayrulewidth}
\multirow{2}{*}{\textbf{Model}} & \multicolumn{5}{c|}{\textbf{ML-1M}}                                                     & \multicolumn{6}{c|}{\textbf{ML-20M}}                                                                      & \multicolumn{5}{c}{\textbf{KuaiRand}}                                                   \\ \cline{2-17} 
                                & \textbf{HR@10}  & \textbf{HR@50}  & \textbf{NG@10}  & \textbf{NG@50}  & \textbf{MRR}    & \textbf{HR@10}  & \textbf{HR@50}  & \textbf{NG@10}  & \textbf{NG@50}  & \textbf{MRR}    & \textbf{Infer.} & \textbf{HR@10}  & \textbf{HR@50}  & \textbf{NG@10}  & \textbf{NG@50}  & \textbf{MRR}    \\ \hline
LinRec                          & 0.2374          & 0.4923          & 0.1301          & 0.1860          & 0.1124          & 0.1629          & 0.3950          & 0.0845          & 0.1350          & 0.0748          & 0.98            & 0.0799          & 0.2454          & 0.0396          & 0.0750          & 0.0382          \\
FLASH                           & 0.2731          & 0.5391          & 0.1506          & 0.2096          & 0.1288          & 0.2484          & 0.5028          & 0.1368          & 0.1928          & 0.1179          & 1.26            & 0.0977          & 0.2844          & 0.0495          & 0.0896          & 0.0468          \\
GRU4Rec                         & 0.2840          & 0.5460          & 0.1551          & 0.2130          & 0.1311          & 0.2411          & 0.4987          & 0.1311          & 0.1877          & 0.1128          & 1.00            & 0.0954          & 0.2686          & 0.0482          & 0.0854          & 0.0452          \\
SASRec                          & 0.2832          & 0.5478          & 0.1583          & 0.2170          & 0.1360          & 0.2645          & 0.5225          & 0.1465          & 0.2033          & 0.1257          & 1.05            & 0.1003          & 0.2878          & 0.0505          & 0.0909          & 0.0475          \\
LRURec                          & 0.2910          & 0.5550          & 0.1641          & 0.2227          & 0.1409          & 0.2732          & 0.5326          & 0.1514          & 0.2086          & 0.1295          & 1.21            & 0.0825          & 0.2391          & 0.0415          & 0.0751          & 0.0394          \\
Mamba4Rec                       & 0.3140          & 0.5775          & 0.1766          & 0.2351          & 0.1500          & 0.2889          & 0.5491          & 0.1619          & 0.2194          & 0.1384          & 1.35            & 0.0903          & 0.2560          & 0.0455          & 0.0810          & 0.0427          \\
LLaMa                           & 0.3075          & 0.5799          & 0.1704          & 0.2305          & 0.1441          & 0.3045          & 0.5685          & 0.1723          & 0.2306          & 0.1472          & 1.26            & 0.1071          & 0.3060          & 0.0542          & 0.0971          & 0.0508          \\
HSTU                            & 0.2955          & 0.5727          & 0.1667          & 0.2279          & 0.1433          & 0.2929          & 0.5582          & 0.1653          & 0.2239          & 0.1418          & 1.27            & 0.1050          & 0.2911          & 0.0535          & 0.0936          & 0.0498          \\
FuXi-$\alpha$                          & 0.3205          & 0.5892          & 0.1814          & 0.2409          & 0.1545          & 0.3359          & 0.5971          & 0.1956          & 0.2534          & 0.1677          & 1.34            & 0.1108          & 0.3081          & 0.0555          & 0.0981          & 0.0513          \\
FuXi-$\beta$                          & 0.3207          & 0.5811          & 0.1864          & 0.2443          & 0.1605          & 0.3324          & 0.5909          & 0.1945          & 0.2517          & 0.1673          & 1.03            & 0.1064          & 0.2979          & 0.0534          & 0.0946          & 0.0495          \\
\textbf{FuXi-$\gamma$}                 & 0.3255          & 0.5928          & 0.1850          & 0.2443          & 0.1576          & 0.3346          & 0.5918          & 0.1960          & 0.2529          & 0.1684          & 1.00            & \underline{0.1118}          & \underline{0.3082}          & \underline{0.0562}          & \underline{0.0984}          & \underline{0.0518}          \\ \hline
LinRec-Large                    & 0.0385          & 0.1301          & 0.0189          & 0.0381          & 0.0200          & 0.0483          & 0.1276          & 0.0235          & 0.0404          & 0.0221          & 0.94            & 0.0503          & 0.1153          & 0.0247          & 0.0386          & 0.0223          \\
FLASH-Large                     & 0.2807          & 0.5464          & 0.1552          & 0.2140          & 0.1324          & 0.2908          & 0.5470          & 0.1639          & 0.2205          & 0.1402          & 1.78            & 0.0983          & 0.2852          & 0.0492          & 0.0894          & 0.0464          \\
GRU4Rec-Large                   & 0.2285          & 0.4857          & 0.1226          & 0.1790          & 0.1054          & 0.2001          & 0.4524          & 0.1048          & 0.1601          & 0.0910          & 1.07            & 0.0885          & 0.2463          & 0.0446          & 0.0786          & 0.0419          \\
SASRec-Large                    & 0.0375          & 0.1303          & 0.0177          & 0.0371          & 0.0186          & 0.0414          & 0.1218          & 0.0207          & 0.0382          & 0.0208          & 1.12            & 0.0470          & 0.1054          & 0.0251          & 0.0376          & 0.0233          \\
LRURec-Large                    & 0.0392          & 0.1325          & 0.0188          & 0.0384          & 0.0195          & 0.0454          & 0.1288          & 0.0225          & 0.0404          & 0.0220          & 1.64            & 0.0525          & 0.1201          & 0.0269          & 0.0410          & 0.0242          \\
Mamba4Rec-Large                 & 0.0505          & 0.1563          & 0.0253          & 0.0477          & 0.0254          & 0.0441          & 0.1367          & 0.0220          & 0.0418          & 0.0220          & 1.64            & 0.0513          & 0.1361          & 0.0268          & 0.0448          & 0.0254          \\
LLaMa-Large                     & 0.3259          & 0.5920          & 0.1856          & 0.2450          & 0.1584          & 0.3459          & 0.6069          & 0.2020          & 0.2598          & 0.1730          & 1.78            & 0.1073          & 0.3042          & 0.0540          & 0.0964          & 0.0503          \\
HSTU-Large                      & 0.3293          & 0.5930          & 0.1867          & 0.2454          & 0.1586          & 0.3405          & 0.6007          & 0.1992          & 0.2568          & 0.1710          & 1.79            & 0.1065          & 0.2944          & 0.0535          & 0.0938          & 0.0493          \\
FuXi-$\alpha$-Large                    & 0.3321          & 0.5908          & 0.1919          & 0.2493          & 0.1641          & 0.3535          & 0.6086          & 0.2095          & 0.2660          & 0.1801          & 2.03            & 0.1107          & 0.3072          & 0.0552          & 0.0974          & 0.0508          \\
FuXi-$\beta$-Large                    & \underline{0.3390}          & \underline{0.6020}          & \underline{0.1945}          & \underline{0.2529}          & \underline{0.1656}          & \underline{0.3551}          & \underline{0.6106}          & \underline{0.2109}          & \underline{0.2675}          & \underline{0.1814}          & 1.15            & 0.1079          & 0.3046          & 0.0543          & 0.0966          & 0.0505          \\
\textbf{FuXi-$\gamma$-Large}           & \textbf{0.3423*} & \textbf{0.6029*} & \textbf{0.1975*} & \textbf{0.2552*} & \textbf{0.1682*} & \textbf{0.3588*} & \textbf{0.6137*} & \textbf{0.2135*} & \textbf{0.2700*} & \textbf{0.1836*} & 1.00            & \textbf{0.1137*} & \textbf{0.3162*} & \textbf{0.0571*} & \textbf{0.1006*} & \textbf{0.0525*} \\
\Xhline{2\arrayrulewidth}
\end{tabular}
}
\end{table*}

\subsubsection{Datasets.}
To evaluate the performance of our FuXi-$\gamma$, we conduct experiments on four real-world datasets, including three public datasets and one large-scale industrial dataset. 
These datasets involve three different scenarios, including movie, video, and music. 
For dataset preprocessing, we follow the common practice in~\cite{zhai2024actions, ye2025fuxi, ye2025beta}. 
The statistics of the processed datasets are shown in Table~\ref{table: statistics of the datasets}.

\begin{itemize}[leftmargin=1.2em]

\item {\textbf{MovieLens}~\cite{harper2015movielens}}: 
This is a widely used benchmark dataset for evaluating recommendation algorithms. 
The dataset includes users’ rating and tagging activities. 
In this paper, we adopt two well-established versions for our experiments, i.e., MovieLens-1M (\textbf{ML-1M}\footnote{\url{https://grouplens.org/datasets/movielens/1m/}}) and MovieLens-20M (\textbf{ML-20M}\footnote{\url{https://grouplens.org/datasets/movielens/20m/}}).

\item {\textbf{KuaiRand}~\cite{gao2022kuairand}}: 
The dataset\footnote{\url{https://kuairand.com/}} is collected from sequential interaction logs of Kuaishou, a prominent short-video sharing mobile application. 
This platform shows high engagement, averaging 50+ interactions per active user. 

\item {\textbf{Industrial}}: 
This dataset is constructed from the user records of an industrial mainstream music app, which has tens of millions of active users every month. 
This complex dataset can better evaluate the robustness and effectiveness of models. 

\end{itemize}

\subsubsection{Baselines.}
For a competitive comparison, we evaluate FuXi-$\gamma$ against a broad set of representative baselines spanning diverse architectures, including LinRec~\cite{liu2023linrec}, FLASH~\cite{hua2022transformer}, GRU4Rec~\cite{hidasi2015session}, SASRec~\cite{kang2018self}, LRURec~\cite{yue2024linear}, Mamba4Rec~\cite{liu2024mamba4rec}, LLaMa~\cite{grattafiori2024llama}, HSTU~\cite{zhai2024actions}, FuXi-$\alpha$~\cite{ye2025fuxi}, and FuXi-$\beta$~\cite{ye2025beta}. 
These baselines collectively cover the key and efficient paradigms in modern sequential recommendation, such as RNN-based, Transformer-based, and Mamba-based architectures, ensuring a fair and comprehensive evaluation.

\subsubsection{Evaluation Metrics.}
We employ three well-established evaluation metrics to evaluate recommendation performance: Hit Ratio (HR), Normalized Discounted Cumulative Gain (NDCG), and Mean Reciprocal Rank (MRR)~\cite{zhai2024actions, ye2025fuxi}. 
HR@K measures whether the ground-truth item appears within the top-K positions of the recommendation list. 
NDCG@K evaluates top-K recommendation quality by assigning higher scores to relevant items ranked closer to the top. 
MRR evaluates the ranking quality by computing the reciprocal rank of the first relevant item in the recommendation results. 
For all these metrics, higher values indicate better recommendation performance. 
Following common practice~\cite{ye2025fuxi, ye2025beta}, we report HR@K and NDCG@K with K = 10, 50 by default.

\subsubsection{Implementation Details.}
\label{sec: implementation details}
We implement FuXi-$\gamma$ using PyTorch\footnote{\url{https://pytorch.org/}}. 
For fair comparison, all models share the same hyper-parameter settings, primarily following FuXi-$\alpha$~\cite{ye2025fuxi} and FuXi-$\beta$~\cite{ye2025beta}. 
Specifically, we use the AdamW optimizer~\cite{loshchilov2017decoupled} with a learning rate of 0.001 and a batch size of 128. 
The hidden dimensions are set to 50, 64, 256, and 256 for ML-1M, KuaiRand, ML-20M, and Industrial datasets, respectively. 
The dropout rate is set to 0.2, and the number of negative samples is set to 128. 
For our FuXi-$\gamma$, the temporal encoder’s decay parameter $\gamma$ is set to 0.8 for movie and video scenarios, and 0.9 for music. 
This setting reflects the intuition that music consumption typically exhibits stronger long-term user preferences, thus requiring a slower decay rate to preserve long-range interest signals, whereas movie/video interactions are more sensitive to recent behaviors. 
To evaluate base modeling capacity, we set the number of layers to 2. 
To analyze scaling effects, we extend the models to 8 layers (i.e., 4$\times$ deeper) and denote the variants as "XX-Large". 
For the Industrial dataset, we adopt 4 layers, consistent with the deployed online model. 
To enable efficient large-scale training, we utilize multi-NPU parallelism via the Accelerate library~\cite{kotter2012accelerate}.

\subsection{Overall Performance}

\subsubsection{Recommendation Performance on Public Datasets.}
As shown in Table~\ref{table: overall recommendation performance comparison on the public datasets}, we summarize the key observations as follows: 

\begin{itemize}[leftmargin=1.2em]

\item
Our proposed FuXi-$\gamma$ consistently outperforms state-of-the-art baselines across all datasets. 
Under the 8-layer configuration, FuXi-$\gamma$ surpasses other autoregressive models by an average of 3.79\% in HR@10, 2.37\% in HR@50, 4.46\% in NDCG@10, 3.49\% in NDCG@50, and 4.24\% in MRR. 
Even with shallow 2-layer configuration, FuXi-$\gamma$ still achieves the best or second-best performance, demonstrating its strong ability to capture user interest patterns. 
Notably, our proposed exponential-power temporal encoder contributes meaningfully to this improvement. 
An in-depth study of this module is provided in Section~\ref{sec: in-depth study of temporal encoder}. 

\item
Generative models based on autoregressive architectures overall outperform the traditional methods, which aligns with previous findings~\cite{zhai2024actions, guo2024scaling, ye2025fuxi, xie2025breaking, ye2025beta}. 
For example, increasing the depth of GRU4Rec or SASRec from 2 to 8 layers results in performance degradation, indicating that traditional architectures struggle to model complex item dependencies. 
In contrast, generative models exhibit strong scaling effects, highlighting their potential for more expressive sequence recommendation modeling. 

\item
Our FuXi-$\gamma$ achieves excellent inference efficiency, surpassed only by LinRec on ML-20M. 
The efficiency gains primarily stem from three factors: 
(1) Nearly pure matrix-based architecture: Each FuXi-$\gamma$ block consists almost entirely of hardware-friendly matrix operations, enabling high hardware utilization. 
(2) Minimalistic single-head attention: FuXi-$\gamma$ employs no query-key attention and no multi-head expansion, avoiding the costly projections, scaling operations, activations, head-wise computations, and tensor-splitting overheads typical of Transformers. 
(3) Efficient temporal encoder: The exponential-power temporal encoder relies on continuous matrix operations rather than discrete bucket lookups, eliminating irregular memory access. 
Further evaluation of efficiency against strong generative baselines across varying sequence lengths is provided in Section~\ref{sec: efficiency performance}.

\end{itemize}

\begin{table}[H]
\centering
\caption{Overall recommendation performance comparison on the industrial dataset. The best result is highlighted in bold, and the second-best result is underlined. Superscript * indicates the statistically significant improvements (i.e., two-sided t-test with $p$ < 0.05).}
\label{table: overall recommendation performance comparison on the industrial dataset}
\scalebox{0.9}{
\begin{tabular}{c|ccccc}
\Xhline{2\arrayrulewidth}
\multirow{2}{*}{\textbf{Model}} & \multicolumn{5}{c}{\textbf{Industrial}}                                                          \\ \cline{2-6} 
                       & \textbf{HR@10}           & \textbf{HR@50}           & \textbf{NG@10}           & \textbf{NG@50}           & \textbf{MRR}             \\ \hline
SASRec                 & 0.2954          & 0.5742          & 0.1633          & 0.2256          & 0.1397          \\
LLaMa                  & 0.3714          & 0.6231          & 0.2216          & 0.2779          & 0.1907          \\
HSTU                   & 0.3831          & 0.6336          & 0.2294          & 0.2855          & 0.1972          \\
FuXi-$\alpha$                 & 0.4060          & 0.6483          & 0.2479          & 0.3022          & 0.2137          \\
FuXi-$\beta$                 & \underline{0.4734}          & \underline{0.6776}          & \underline{0.3174}          & \underline{0.3633}          & \underline{0.2811}          \\
\textbf{FuXi-$\gamma$}        & \textbf{0.5064*} & \textbf{0.6894*} & \textbf{0.3559*} & \textbf{0.3969*} & \textbf{0.3196*} \\ \Xhline{2\arrayrulewidth}
\end{tabular}
}
\end{table}

\subsubsection{Recommendation Performance on Industrial Dataset.}

As shown in Table~\ref{table: overall recommendation performance comparison on the industrial dataset}, FuXi-$\gamma$ shows more significant advantages on our large-scale industrial music dataset. 
Compared with other autoregressive baselines, it achieves \textbf{25.06\%} higher HR@10 and \textbf{42.86\%} higher NDCG@10 on average. 
Against the strongest baseline FuXi-$\beta$, our FuXi-$\gamma$ achieves gains of 6.97\% in HR@10, 1.74\% in HR@50, 12.13\% in NDCG@10, 9.25\% in NDCG@50, and 13.70\% in MRR. 
These results further demonstrate that FuXi-$\gamma$ generalizes well in various application scenarios, benefiting from our temporal encoder that adaptively models both long-term and short-term user interests.

\subsubsection{Ablation Study.}

Table~\ref{table: ablation study} reports the results of ablation studies on the three public datasets. 
Among all components, our proposed exponential-power temporal encoder proves to be the most critical. 
It allows the model to assign differentiated weights to items based on time intervals, effectively capturing the temporal evolution of user preferences. 
The positional encoder contributes moderately. 
Since item representations already incorporate absolute positional embeddings, and temporal information implicitly encodes sequential order, the relative positional channel provides marginal yet complementary gains. 
The SwiGLU FFN’s impact is relatively minor, suggesting that FuXi-$\gamma$’s dual-channel self-attention layer sufficiently models sequence dependencies. 
Nevertheless, its implicit interaction provides a minor boost to model expressiveness.

\begin{table}[H]
\centering
\caption{Ablation study based on FuXi-$\gamma$.}
\label{table: ablation study}
\scalebox{0.74}{
\setlength{\tabcolsep}{3pt}
\begin{tabular}{c|cc|cc|cc}
\Xhline{2\arrayrulewidth}
\multirow{2}{*}{\textbf{Setting}} & \multicolumn{2}{c|}{\textbf{ML-1M}} & \multicolumn{2}{c|}{\textbf{ML-20M}} & \multicolumn{2}{c}{\textbf{KuaiRand}} \\ \cline{2-7} 
                                  & \textbf{HR@10}   & \textbf{NG@10}   & \textbf{HR@10}    & \textbf{NG@10}   & \textbf{HR@10}    & \textbf{NG@10}    \\ \hline
\textbf{FuXi-$\gamma$}            & \textbf{0.3423}  & \textbf{0.1975}  & \textbf{0.3588}   & \textbf{0.2135}  & \textbf{0.1137}   & \textbf{0.0571}            \\
w/o SwiGLU FFN                    & 0.3361           & 0.1931           & 0.3496            & 0.2059           & 0.1117            & 0.0570            \\
w/o Positional Encoder            & 0.3281           & 0.1865           & 0.3548            & 0.2096           & 0.1113            & 0.0559            \\ 
w/o Temporal Encoder              & 0.3120           & 0.1767           & 0.3248            & 0.1882           & 0.1043            & 0.0521            \\ \Xhline{2\arrayrulewidth}
\end{tabular}
}
\end{table}

\subsubsection{Efficiency Performance.}
\label{sec: efficiency performance}

\begin{table*}[]
\centering
\caption{Compatibility comparison of temporal encoders.}
\label{table: effectiveness comparison of temporal encoder}
\scalebox{0.95}{
\begin{tabular}{cc|ccc|ccc|ccc}
\Xhline{2\arrayrulewidth}
                                        &                                             & \multicolumn{3}{c|}{\textbf{ML-1M}}                                                                                         & \multicolumn{3}{c|}{\textbf{ML-20M}}                                                                                        & \multicolumn{3}{c}{\textbf{KuaiRand}}                                                                                       \\ \cline{3-11} 
\multirow{-2}{*}{\textbf{Architecture}} & \multirow{-2}{*}{\textbf{Temporal Encoder}} & \textbf{HR@10}                          & \textbf{NG@10}                          & \textbf{MRR}                            & \textbf{HR@10}                          & \textbf{NG@10}                          & \textbf{MRR}                            & \textbf{HR@10}                          & \textbf{NG@10}                          & \textbf{MRR}                            \\ \hline
                                        & w/o                                         & 0.3117                                  & 0.1719                                  & 0.1444                                  & 0.3151                                  & 0.1805                                  & 0.1544                                  & 0.1004                                  & 0.0502                                  & 0.0465                                  \\
                                        & Used in HSTU, FuXi-$\alpha$                        & 0.3293                                  & 0.1867                                  & 0.1586                                  & 0.3405                                  & 0.1992                                  & 0.1710                                  & 0.1065                                  & 0.0535                                  & 0.0493                                  \\
                                        & Proposed in FuXi-$\beta$                          & 0.3224                                  & 0.1853                                  & 0.1589                                  & 0.3410                                  & 0.1995                                  & 0.1713                                  & 0.1048                                  & 0.0530                                  & 0.0492                                  \\
\multirow{-4}{*}{HSTU}                  & \cellcolor[HTML]{E0E0E0}Ours                & \cellcolor[HTML]{E0E0E0}\textbf{0.3312}          & \cellcolor[HTML]{E0E0E0}\textbf{0.1897}          & \cellcolor[HTML]{E0E0E0}\textbf{0.1618}          & \cellcolor[HTML]{E0E0E0}\textbf{0.3426}          & \cellcolor[HTML]{E0E0E0}\textbf{0.2012}          & \cellcolor[HTML]{E0E0E0}\textbf{0.1729}          & \cellcolor[HTML]{E0E0E0}\textbf{0.1084}          & \cellcolor[HTML]{E0E0E0}\textbf{0.0551}          & \cellcolor[HTML]{E0E0E0}\textbf{0.0513}          \\ \hline
                                        & w/o                                         & 0.3104                                  & 0.1738                                  & 0.1476                                  & 0.3226                                  & 0.1870                                  & 0.1602                                  & 0.1020                                  & 0.0515                                  & 0.0483                                  \\
                                        & Used in HSTU, FuXi-$\alpha$                        & 0.3321                                  & 0.1919                                  & 0.1641                                  & 0.3535                                  & 0.2095                                  & 0.1801                                  & 0.1107                                  & 0.0552                                  & 0.0508                                  \\
                                        & Proposed in FuXi-$\beta$                          & 0.3338                                  & 0.1891                                  & 0.1599                                  & 0.3538                                  & 0.2098                                  & 0.1803                                  & 0.1071                                  & 0.0539                                  & 0.0500                                  \\
\multirow{-4}{*}{FuXi-$\alpha$}                & \cellcolor[HTML]{E0E0E0}Ours                & \cellcolor[HTML]{E0E0E0}\textbf{0.3361}          & \cellcolor[HTML]{E0E0E0}\textbf{0.1940}          & \cellcolor[HTML]{E0E0E0}\textbf{0.1657}          & \cellcolor[HTML]{E0E0E0}\textbf{0.3540}          & \cellcolor[HTML]{E0E0E0}\textbf{0.2104}          & \cellcolor[HTML]{E0E0E0}\textbf{0.1810}          & \cellcolor[HTML]{E0E0E0}\textbf{0.1114}          & \cellcolor[HTML]{E0E0E0}\textbf{0.0566}          & \cellcolor[HTML]{E0E0E0}\textbf{0.0520}          \\ \hline
                                        & w/o                                         & 0.3120                                  & 0.1767                                  & 0.1508                                  & 0.3248                                  & 0.1882                                  & 0.1611                                  & 0.1043                                  & 0.0521                                  & 0.0488                                  \\
                                        & Used in HSTU, FuXi-$\alpha$                        & 0.3350                                  & 0.1931                                  & 0.1655                                  & 0.3550                                  & 0.2107                                  & 0.1811                                  & 0.1111                                  & 0.0557                                  & 0.0512                                  \\
                                        & Proposed in FuXi-$\beta$                          & 0.3390                                  & 0.1945                                  & 0.1656                                  & 0.3551                                  & 0.2109                                  & 0.1814                                  & 0.1079                                  & 0.0543                                  & 0.0505                                  \\
\multirow{-4}{*}{FuXi-$\gamma$, FuXi-$\beta$}      & \cellcolor[HTML]{E0E0E0}Ours                & \cellcolor[HTML]{E0E0E0}\textbf{0.3423} & \cellcolor[HTML]{E0E0E0}\textbf{0.1975} & \cellcolor[HTML]{E0E0E0}\textbf{0.1682} & \cellcolor[HTML]{E0E0E0}\textbf{0.3588} & \cellcolor[HTML]{E0E0E0}\textbf{0.2135} & \cellcolor[HTML]{E0E0E0}\textbf{0.1836} & \cellcolor[HTML]{E0E0E0}\textbf{0.1137} & \cellcolor[HTML]{E0E0E0}\textbf{0.0571} & \cellcolor[HTML]{E0E0E0}\textbf{0.0525} \\ \Xhline{2\arrayrulewidth}
\end{tabular}
}
\end{table*}

\begin{figure}[H]
  \centering
  \subfigure[Training Efficiency]{\label{fig: kuairand_train}\includegraphics[width=0.48\linewidth]{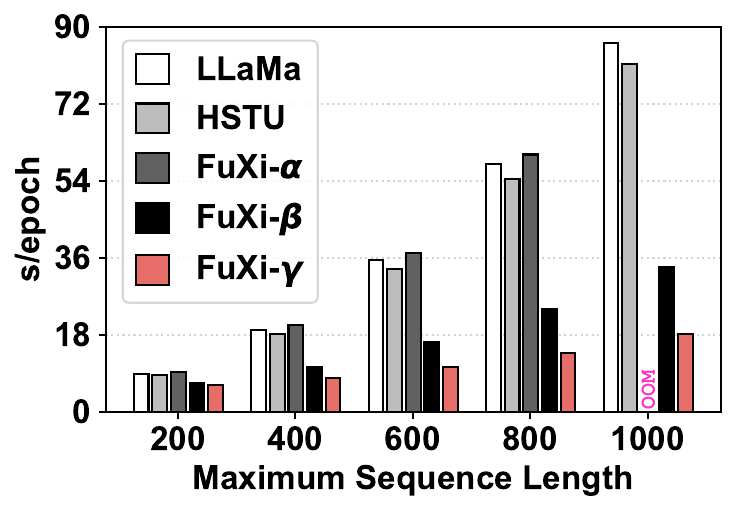}}\hspace{2mm}
  \subfigure[Inference Efficiency]{\label{fig: kuairand_inference}\includegraphics[width=0.48\linewidth]{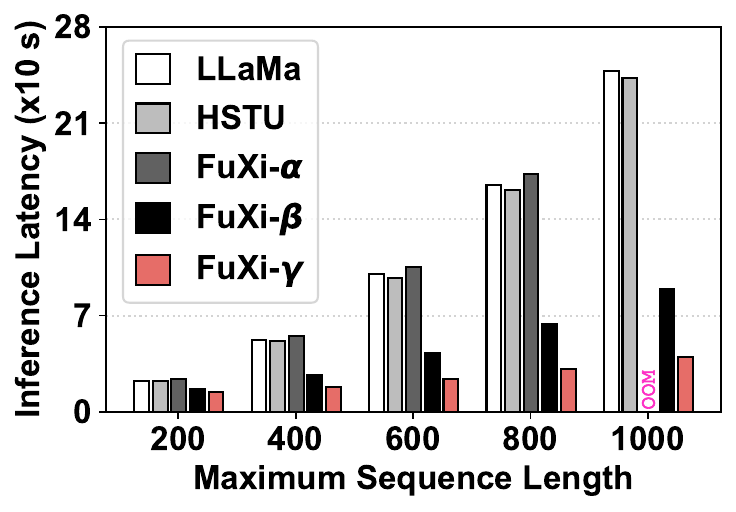}}
  \caption{Overall efficiency performance comparison.}
  \label{fig: overall efficiency}
\end{figure}

Having established the effectiveness of FuXi-$\gamma$, we now evaluate its computational efficiency across varying sequence lengths. 
Comparisons are made against other autoregressive generative models, which uniformly outperform traditional baselines in both accuracy and scaling effects. 
We report results on KuaiRand, with consistent trends observed across other datasets. 
(1) Training Efficiency: 
As shown in Figure~\ref{fig: kuairand_train}, FuXi-$\gamma$ stably achieves the highest training efficiency across all sequence lengths, with its advantage growing more pronounced at longer sequences. 
At a length of 1000, it achieves speedups of \textbf{4.74$\times$}, 4.48$\times$, and 1.86$\times$ over LLaMa, HSTU, and FuXi-$\beta$, respectively. 
Notably, FuXi-$\alpha$ encounters out-of-memory (OOM) issue at this scale, due to its computationally intensive three-channel architecture. 
(2) Inference Efficiency: 
Figure~\ref{fig: kuairand_inference} further demonstrates FuXi-$\gamma$'s leading inference efficiency, with the performance gap similarly widening at longer sequence lengths. 
At length 1000, FuXi-$\gamma$ achieves inference speedups of \textbf{6.18$\times$}, 6.07$\times$, and 2.24$\times$ over LLaMa, HSTU, and FuXi-$\beta$, respectively. 
These results underscore FuXi-$\gamma$’s strong scalability and practical suitability for long-sequence recommendation, attributed to its streamlined dual-channel architecture and lightweight exponential-power temporal encoder.

\subsection{Study of Temporal Encoder}
\label{sec: in-depth study of temporal encoder}

\begin{figure}
  \centering
  \subfigure[Under Different Max. Seq. Lengths]{\label{fig: tp_efficiency_seqlen}\includegraphics[width=0.48\linewidth]{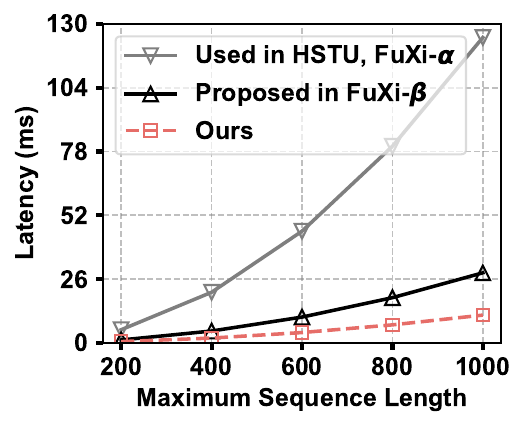}}\hspace{2mm}
  \subfigure[Under Different Batch Sizes]{\label{fig: tp_efficiency_batchsize}\includegraphics[width=0.48\linewidth]{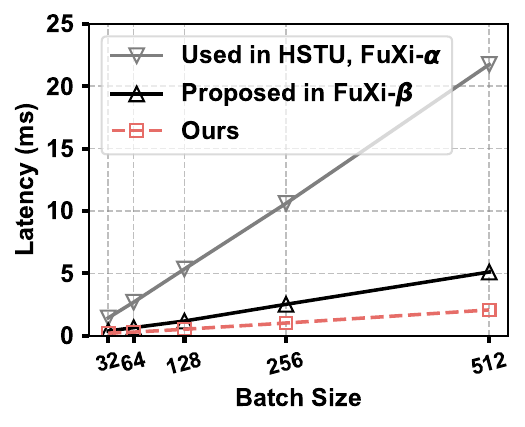}}
  \caption{Efficiency comparison of temporal encoders.}
  \label{fig: efficiency comparison of temporal encoder}
\end{figure}

\subsubsection{Compatibility.}

As shown in Table~\ref{table: effectiveness comparison of temporal encoder}, we evaluate the compatibility of different temporal encoders by integrating them into three representative state-of-the-art architectures. 
Our exponential-power temporal encoder robustly yields the highest accuracy improvements across all architectures on all datasets. 
On average, it improves HR@10, NDCG@10, and MRR by 8.82\%, \textbf{11.16\%}, and 11.15\%, respectively, compared to the non-temporal baseline. 
This performance gain stems from its alignment with human memory decay patterns via exponential modeling. 
In addition, the temporal encoders used in HSTU, FuXi-$\alpha$, and FuXi-$\beta$ also lead to varying degrees of improvement, reaffirming the critical role of temporal modeling in sequential recommendation task.

\subsubsection{Efficiency.}

Figure~\ref{fig: efficiency comparison of temporal encoder} presents a comparative analysis of the computational efficiency of our exponential-power temporal encoder against two strong baselines. 
Across all sequence lengths, our encoder achieves the lowest latency, owing to its continuous memory access pattern, lightweight computation, and pre-conversion of data type. 
In contrast, bucket-based encoder (e.g., adopted by HSTU and FuXi-$\alpha$) exhibits severe degradation in efficiency as sequence length increases. 
This is primarily due to their discontinuous memory addressing, which is fundamentally incompatible with modern parallel hardware architectures. 
Although FuXi-$\beta$ improves upon this with an inverse-proportional decay function, it still falls short of our method. 
At a sequence length of 1000, our encoder achieves \textbf{11.00$\times$} and 2.52$\times$ speedup over two baselines, respectively. 
Furthermore, while the latency of all encoders scales proportionally with batch size, ours reliably maintains superior efficiency.

\subsubsection{Visualization.}

\begin{figure}
  \centering
  \subfigure[Layer-1 on ML-1M]{\label{fig: tp_wgt_ml1m}\includegraphics[width=0.452\linewidth]{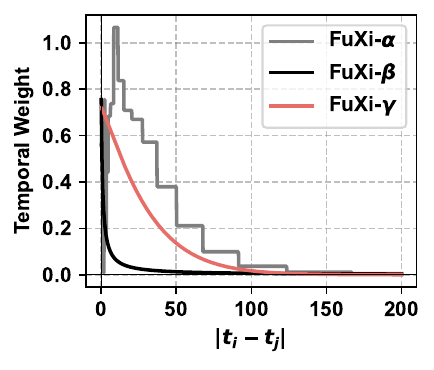}}\hspace{3mm}
  \subfigure[Layer-1 on ML-20M]{\label{fig: tp_wgt_ml20m}\includegraphics[width=0.452\linewidth]{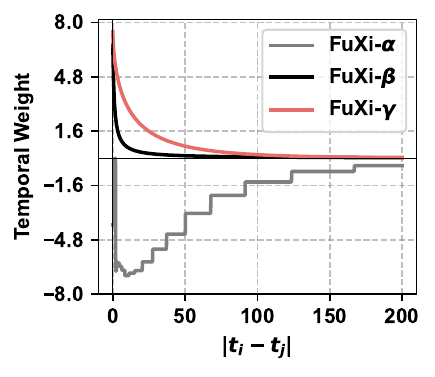}}
  \caption{Visualization comparison of temporal encoders.}
  \label{fig: visualization comparison of temporal encoder}
\end{figure}

Figure~\ref{fig: visualization comparison of temporal encoder} illustrates the encoding patterns of different temporal encoders. 
The horizontal axis represents the relative time interval between two items, and the vertical axis denotes the corresponding attention weight assigned. 
We visualize the first-layer temporal encoders of FuXi-$\alpha$, FuXi-$\beta$, and FuXi-$\gamma$ trained on ML-1M and ML-20M. 
All encoders follow the intuitive pattern that longer intervals receive lower weights, emphasizing the importance of recent interactions. 
FuXi-$\beta$ exhibits a steep initial decline, allocating excessive weights to very short-term intervals and limiting long-range modeling. 
FuXi-$\alpha$ shows a sawtooth pattern, caused by its bucketization strategy, where multiple time intervals are mapped to the same bucket, and adjacent buckets are not smoothly connected. 
In contrast, our FuXi-$\gamma$ generates a smooth, continuous decay curve, thanks to its exponential-power function design. 
This allows it to differentiate both short- and long-term intervals.

\subsubsection{Impact of Hyper-Parameter $\gamma$.}

\begin{figure}
  \centering
  \subfigure[On ML-20M]{\label{fig: gamma_ml20m}\includegraphics[width=0.48\linewidth]{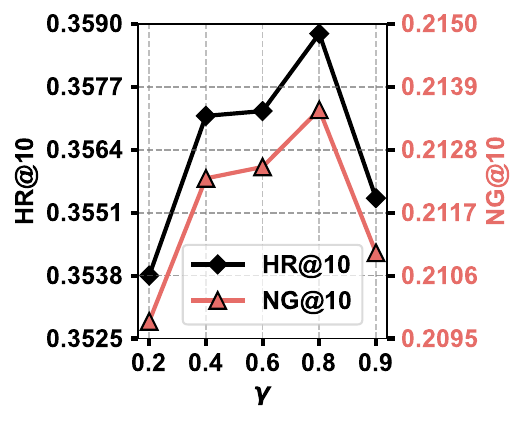}}\hspace{2mm}
  \subfigure[On KuaiRand]{\label{fig: gamma_kuairand}\includegraphics[width=0.48\linewidth]{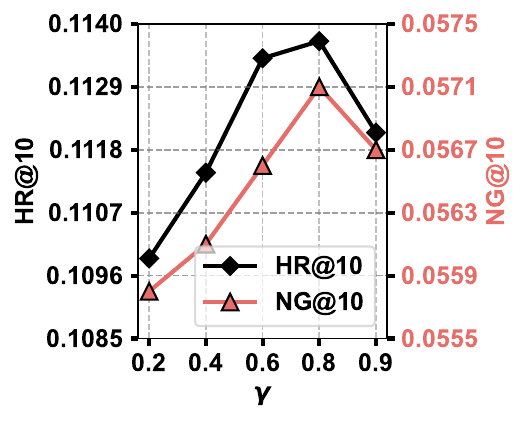}}
  \caption{Hyper-parameter study of $\gamma$.}
  \label{fig: hyper-parameter}
\end{figure}

Figure~\ref{fig: hyper-parameter} presents the impact of the decay-rate hyper-parameter $\gamma$ on model performance in two domains: movie and video. 
As $\gamma$ increases, temporal decay becomes slower. 
We observe that $\gamma=0.8$ achieves the best performance in both domains, balancing the preservation of long-term preferences with sensitivity to short-term variations. 
A too-fast decay (e.g., smaller $\gamma$) weakens long-range signals, while a too-slow decay (e.g., larger $\gamma$) dilutes recent behavior effects. 
As mentioned in Section~\ref{sec: implementation details}, we achieve the best results on our industrial music dataset with $\gamma=0.9$, aligning with the domain characteristic that music preferences are more persistent and long-term oriented.

\begin{figure}
  \centering
  \subfigure[NDCG@10 on ML-20M]{\label{fig: sparse_ng_ml20m}\includegraphics[width=0.48\linewidth]{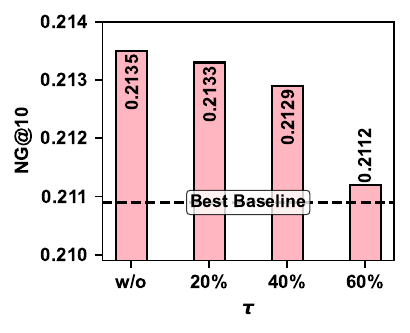}}\hspace{2mm}
  \subfigure[FLOPs on ML-20M]{\label{fig: sparse_flops_ml20m}\includegraphics[width=0.48\linewidth]{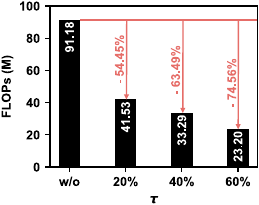}}
  \caption{Impact of configured pruning ratio $\tau$ on ML-20M.}
  \label{fig: sparse-ratio-ml20m}
\end{figure}

\subsection{Study of Sparse Positional Mechanism}

Although FuXi-$\gamma$ already exhibits strong efficiency, we further investigate deployment-oriented optimizations by introducing the diagonal-sparse positional mechanism. 
We evaluate its effects on both effectiveness and efficiency across three representative scenarios: ML-20M (movie), KuaiRand (video), and Industrial (music).

\subsubsection{Impact of Configured Pruning Ratio $\tau$.}
\label{sec: Impact of Configured Pruning Ratio}

\begin{figure}
  \centering
  \subfigure[NDCG@10 on KuaiRand]{\label{fig: sparse_ng_kuairand}\includegraphics[width=0.48\linewidth]{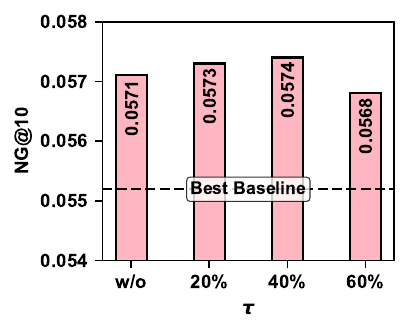}}\hspace{2mm}
  \subfigure[FLOPs on KuaiRand]{\label{fig: sparse_flops_kuairand}\includegraphics[width=0.48\linewidth]{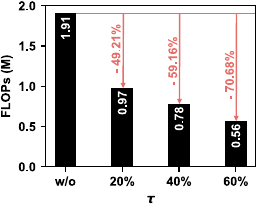}}
  \caption{Impact of configured pruning ratio $\tau$ on KuaiRand.}
  \label{fig: sparse-ratio-kuairand}
\end{figure}

As shown in Figure~\ref{fig: sparse-ratio-ml20m}, on ML-20M, increasing the pruning ratio $\tau$ leads to a mild and acceptable accuracy drop. 
Even at $\tau = 60\%$, FuXi-$\gamma$ retains 98.92\% of its original accuracy, still outperforming the strongest competing baseline. 
Meanwhile, the FLOPs of positional attention are reduced by \textbf{74.56\%}, highlighting substantial computational savings. 
As shown in Figure~\ref{fig: sparse-ratio-kuairand}, on KuaiRand, we observe a slight accuracy improvement at moderate sparsity levels, suggesting that pruning redundant attention blocks can even enhance generalization. 
At $\tau = 60\%$, NDCG@10 decreases by only 0.53\%, while FLOPs are reduced by 70.68\%. 
As shown in Figure~\ref{fig: sparse-ratio-industrial} in Appendix~\ref{appendix: pruning on industrial}, on Industrial, the sparsity mechanism incurs almost no accuracy loss, substantially improving its deployability. 
These results confirm that our semi-structured pruning method effectively eliminates redundancy in positional attention while retaining critical information.

\subsubsection{Impact of Stride Size $s$.}

Table~\ref{table: stride sizes} summarizes the effect of varying stride size $s$ on accuracy and parameter density, using ML-20M as a representative dataset. 
The results show that $s = 8$ offers the best trade-off, achieving strong accuracy with relatively low parameter density. 
Smaller $s$ leads to finer-grained block division, enabling more precise sparsity, but it narrows the scoring range and limits parallelism due to fragmented block structures. 
In contrast, larger $s$ results in coarse-grained blocks, which reduces the effectiveness of sparsity and may fail to capture fine-grained redundancy. 
Thus, the stride size should be adaptively selected based on the deployment context to balance accuracy, density, and efficiency.

\begin{table}[]
\centering
\caption{Impact of stride size $s$ on ML-20M.}
\label{table: stride sizes}
\setlength{\tabcolsep}{8pt}
\begin{tabular}{@{}ccccc@{}}
\Xhline{2\arrayrulewidth}
Stride  & $s$=4     & $s$=8     & $s$=16    & $s$=32    \\ \hline
HR@10   & 0.3574  & 0.3576  & 0.3556  & 0.3567  \\
NG@10   & 0.2126  & 0.2129  & 0.2115  & 0.2119  \\
Density & 33.14\% & 33.29\% & 33.47\% & 36.84\% \\ \Xhline{2\arrayrulewidth}
\end{tabular}
\end{table}

\begin{figure}
\centering
  \includegraphics[width=0.8\linewidth]{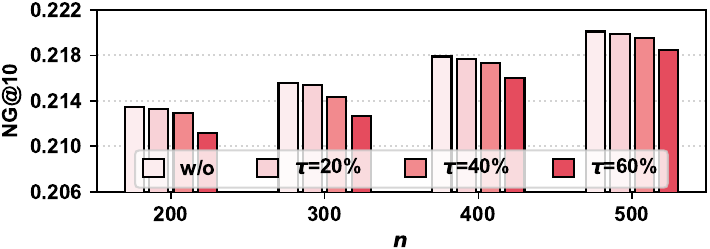}
  \caption{Impact of sequence length $n$ on ML-20M.}
  \label{fig: sparse-seqlen}
\end{figure}

\subsubsection{Impact of Sequence Length $n$.}

Figure~\ref{fig: sparse-seqlen} summarizes the robustness of our sparsity mechanism across different sequence lengths, using ML-20M as a representative dataset. 
The results show that the method remains uniformly stable: even at $\tau = 60\%$, FuXi-$\gamma$ preserves over 98.65\% of its original accuracy, and the performance drop further diminishes as the sequences become longer. 
This trend suggests that longer input sequences may contain higher redundancy in positional interactions, making them more amenable to pruning. 
Consequently, our pruning method holds greater potential for practical deployment in long-sequence recommender systems.

\section{Conclusion}

In this paper, we propose a novel sequential recommendation framework, named FuXi-$\gamma$, to address the dual challenges of efficiency and effectiveness in modeling user interests. 
Our FuXi-$\gamma$ introduces two core innovations: Exponential-Power Temporal Encoder and Diagonal-Sparse Positional Mechanism. 
Exponential-power temporal encoder is inspired by cognitive memory theory and involves only standard matrix operations, enabling it to flexibly capture both short-term and long-term user interests while ensuring efficient execution on modern hardware. 
Diagonal-sparse positional mechanism employs a diagonally semi-structured manner to effectively identify and prune redundant attention blocks, thereby reducing computational cost while preserving recommendation quality. 
Extensive experiments on four real-world datasets demonstrate that our proposed FuXi-$\gamma$ consistently outperforms state-of-the-art baselines in both recommendation accuracy and computational efficiency. 
In future work, we plan to further improve the efficiency and extend the framework to capture more complex user behavior patterns. 

\begin{acks}
This work is supported by the National Natural Science Foundation of China (62372253), the Natural Science Foundation of Tianjin Fund (23JCYBJC00010), and Nankai University School of Optometry and Vision Science Open Fund Program (NKSGP202308). 
The work is also sponsored by Huawei Innovation Research Program. 
\end{acks}

\bibliographystyle{ACM-Reference-Format}
\bibliography{sample-base}

\appendix

\section{Theoretical Discussion of Temporal Encoder}

The exponential-power temporal encoder, formulated as
\begin{equation}
f(x) = \alpha \cdot \gamma^{x^{\beta}},
\label{eq: formulation of temporal encoder}
\end{equation}
extends the classical Ebbinghaus forgetting curve into a generalized exponential kernel. 
Its derivative is
\begin{equation}
f'(x) = \alpha\beta(\ln \gamma)x^{\beta-1}\gamma^{x^{\beta}}.
\label{eq: derivative of temporal encoder}
\end{equation}
When $0 < \beta < 1$, the term $x^{\beta-1} = x^{-(1-\beta)}$ diverges as $x \to 0$, violating Lipschitz continuity and potentially introducing optimization instability. 
To maintain Lipschitz continuity and improve training robustness in this regime, a simple yet effective solution is to shift the input by a small positive constant, replacing $x$ with $x + \epsilon (\epsilon > 0)$.

\section{Analysis of Cold-Start \& Sparse-Case}

Cold-start users and long-tail items are fundamental sparsity challenges in real-world recommendation systems. 
To further assess the robustness of FuXi-$\gamma$, we evaluate model performance under three challenging conditions: 
(1) cold-start users with very limited interaction histories, 
(2) fresh items launched within short time windows, 
and (3) long-tail items with low exposure frequency.

\subsection{Cold-Start Users}

\begin{figure}[H]
  \centering
  \subfigure[HR@10 on Industrial]{\label{fig: user_cold_hr}\includegraphics[width=0.48\linewidth]{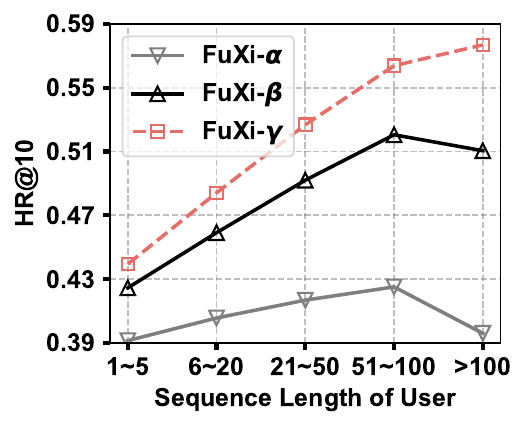}}\hspace{2mm}
  \subfigure[NDCG@10 on Industrial]{\label{fig: user_cold_ng}\includegraphics[width=0.48\linewidth]{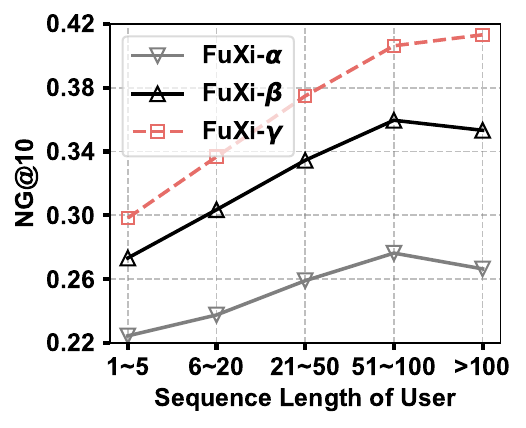}}
  \caption{Analysis of cold-start users on Industrial.}
  \label{fig: cold-start users}
\end{figure}

We partition the industrial dataset into five groups according to users’ historical interaction lengths. 
As shown in Figure~\ref{fig: cold-start users}, our FuXi-$\gamma$ robustly outperforms all baselines across all user groups, including cold-start users with extremely short sequences. 
The performance gap becomes even larger for long-sequence users, indicating that the tunable exponential-power temporal encoder effectively captures both short-term and long-term interest patterns. 

\subsection{Fresh Items}

\begin{figure}[H]
  \centering
  \subfigure[HR@10 on Industrial]{\label{fig: fresh_item_hr}\includegraphics[width=0.48\linewidth]{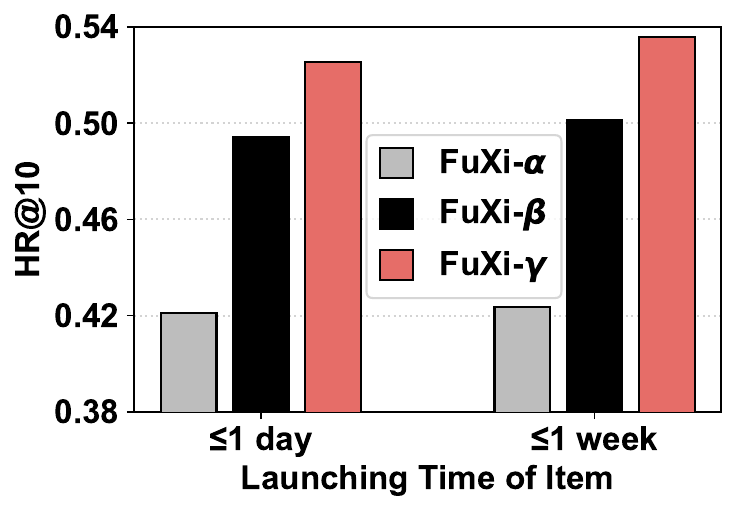}}\hspace{2mm}
  \subfigure[NDCG@10 on Industrial]{\label{fig: fresh_item_ng}\includegraphics[width=0.48\linewidth]{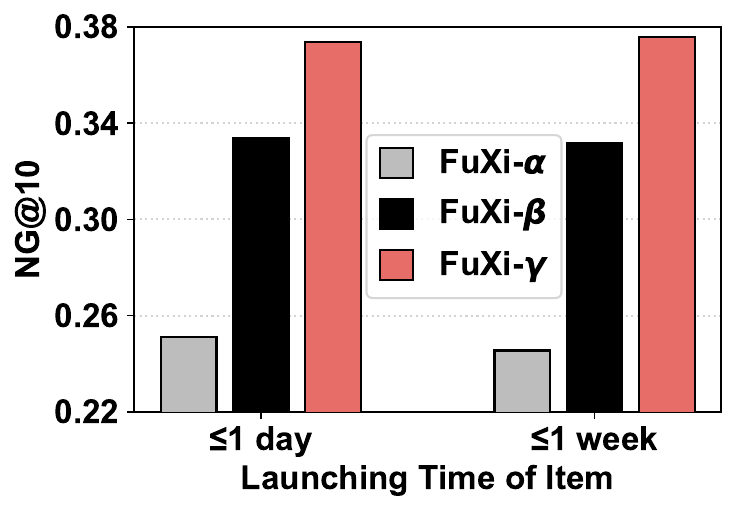}}
  \caption{Analysis of fresh items on Industrial.}
  \label{fig: fresh items}
\end{figure}

We examine the model's ability to recommend fresh items, which lack sufficient historical feedback and are often difficult for models to rank accurately. 
Following the item-launching-time criterion, we evaluate items introduced within $\leq$ 1 day and $\leq$ 1 week. 
As shown in Figure~\ref{fig: fresh items}, FuXi-$\gamma$ achieves the best performance across the two freshness windows. 
These results highlight FuXi-$\gamma$’s strong generalization ability to newly introduced content. 

\subsection{Long-Tail Items}

Following~\cite{zhang2021model, xie2025breaking}, we categorize the top 20\% most interacted items as head and the remainder as long-tail. 
As shown in Table~\ref{table: long-tail items}, FuXi-$\gamma$ stably achieves performance gains over FuXi-$\alpha$ and FuXi-$\beta$ on both subsets. 
We attribute this advantage to FuXi-$\gamma$’s fine-grained temporal encoding, which produces richer sequential representations and alleviates embedding sparsity for long-tail items.

\begin{table}[H]
\centering
\caption{Analysis of long-tail items on ML-20M.}
\label{table: long-tail items}
\scalebox{0.9}{
\begin{tabular}{c|cc|cc}
\Xhline{2\arrayrulewidth}
\multirow{2}{*}{\textbf{Model}} & \multicolumn{2}{c|}{\textbf{Head}} & \multicolumn{2}{c}{\textbf{Tail}} \\ \cline{2-5} 
                       & \textbf{HR@10}       & \textbf{NG@10}       & \textbf{HR@10}       & \textbf{NG@10}      \\ \hline
FuXi-$\alpha$                 & 0.3710      & 0.2205      & 0.0747      & 0.0395     \\
FuXi-$\beta$                 & 0.3705      & 0.2209      & 0.0756      & 0.0409     \\
\textbf{FuXi-$\gamma$}                 & \textbf{0.3749}      & \textbf{0.2234}      & \textbf{0.0789}      & \textbf{0.0413}     \\ \Xhline{2\arrayrulewidth}
\end{tabular}
}
\end{table}

\section{Supplementary Figure}
\label{appendix: pruning on industrial}

\begin{figure}[H]
  \centering
  \subfigure[NDCG@10 on Industrial]{\label{fig: sparse_ng_industrial}\includegraphics[width=0.473\linewidth]{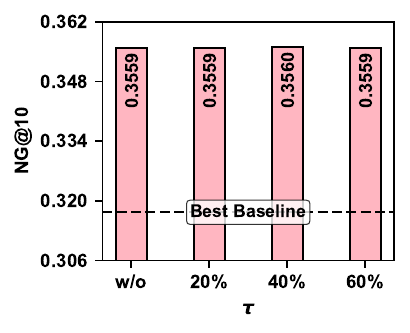}}\hspace{2mm}
  \subfigure[FLOPs on Industrial]{\label{fig: sparse_flops_industrial}\includegraphics[width=0.473\linewidth]{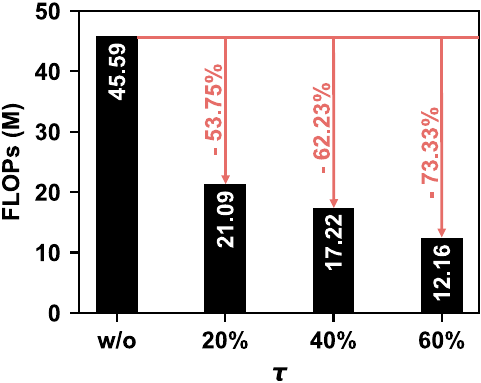}}
  \caption{Impact of configured pruning ratio $\tau$ on Industrial.}
  \label{fig: sparse-ratio-industrial}
\end{figure}

As described in Section~\ref{sec: Impact of Configured Pruning Ratio}, Figure~\ref{fig: sparse-ratio-industrial} demonstrates the impact of configured pruning ratio $\tau$ on Industrial dataset. 

\newpage
\section{Efficiency Discussion about Mamba4Rec}

As shown in Table~\ref{table: overall recommendation performance comparison on the public datasets}, on ML-20M, Mamba4Rec is not universally faster, as linear computational complexity does not guarantee superior runtime in all settings. 
Mamba’s advantage typically emerges for very long sequences (e.g., >2k), while for shorter sequences its efficiency is often lower than that of Transformers~\cite{dao2024transformers, mitra2025characterizing}. 
The reason is that achieving linear scaling relies on recurrent state updates, which introduce non-negligible constant computational overhead. 
In particular, each state update involves complex operations such as matrix multiplications, gating mechanisms, and discretization, requiring additional kernel launches and memory synchronization. 
These operations exhibit limited parallelism, preventing full utilization of AI accelerators. 
Consequently, although Mamba avoids the $O(n^2)$ attention cost, its low hardware parallelism and large constant factors reduce efficiency for typical sequence lengths.

\end{document}